\documentclass[prd,twocolumn,superscriptaddress,altaffilletter,showpacs]{revtex4}
\usepackage{amssymb,amsmath}
\usepackage{comment}
\usepackage{graphicx}

\begin{document}

\title{Mass and spin of Kerr black holes in terms of observational quantities: \\
The dragging effect on the redshift}
\author{Pritam Banerjee}
\email{bpritam@iitk.ac.in}
\affiliation{Department of Physics, Indian Institute of Technology, Kanpur 208016, India.}
\author{Alfredo Herrera--Aguilar}
\email{aherrera@ifuap.buap.mx}
\affiliation{Instituto de F\'{\i}sica, Benem\'erita Universidad Aut\'onoma de Puebla,\\
Apartado Postal J-48, 72570, Puebla, Puebla, M\'exico.}
\author{Mehrab Momennia}
\email{mmomennia@ifuap.buap.mx, momennia1988@gmail.com}
\affiliation{Instituto de F\'{\i}sica, Benem\'erita Universidad Aut\'onoma de Puebla,\\
Apartado Postal J-48, 72570, Puebla, Puebla, M\'exico.}
\author{Ulises Nucamendi}
\email{unucamendi@gmail.com, ulises.nucamendi@umich.mx}
\affiliation{Instituto de F\'{\i}sica y Matem\'{a}ticas, Universidad Michoacana de San
Nicol\'as de Hidalgo,\\
Edificio C--3, Ciudad Universitaria, CP 58040, Morelia, Michoac\'{a}n, M\'{e}%
xico.}
\date{\today }

\begin{abstract}
In this work, we elaborate on the development of a general relativistic
formalism that allows one to analytically express the mass and spin
parameters of the Kerr black hole in terms of observational data: the total
redshift and blueshift of photons emitted by massive geodesic particles
revolving the black hole and their orbital parameters. Thus, we present
concise closed formulas for the mass and spin parameters of the Kerr black
hole in terms of few directly observed quantities in the case of equatorial
circular orbits either when the black hole is static or is moving with
respect to a distant observer.
Furthermore, we incorporate the gravitational dragging effect generated by 
the rotating nature of the Kerr black hole into the analysis and elucidate its 
non-trivial contribution to the expression for the light bending parameter and 
the frequency shifts of photons emitted by orbiting particles that renders 
simple symmetric expressions for the kinematic redshift and blueshift.
We also incorporate the dependency of the frequency shift on the azimuthal
angle, a fact that allows one to express the total redshift/blueshift at
any point of the orbit of the revolving particle for the cases when the
black hole is both static or moving with respect to us. 
These formulas allow one to compute the Kerr black hole parameters by
applying this general relativistic formalism to astrophysical systems like
the megamaser accretion disks orbiting supermassive black holes at the core
of active galactic nuclei. Our results open a new window to implement
parameter estimation studies to constrain black hole variables, and they can
be generalized to black hole solutions beyond Einstein's gravity.

\vskip3mm

\end{abstract}

\maketitle




\section{Introduction and brief overview of the formalism}


Black hole physics has been experiencing a vital boost in the last few years and is
currently a very active research field. On the one hand, the recent
detection of the gravitational waves by the LIGO-Virgo collaborations \cite{LVGW1,LVGW2}
unveiled the existence of black holes through their coalescence. On the
other hand, two research groups of astronomers managed to track the orbital
motion of several stars around the center of our galaxy during the last
three decades and provided convincing dynamical evidence of the existence of
a supermassive black hole hosted at the center of the Milky Way \cite{Ghez,Morris,Eckart,Gillessen}.
These great efforts have been crowned with the 2017 and 2020 Nobel prize in
physics, respectively. Besides, the Event Horizon Telescope Collaboration
imaged the shadow of a supermassive black hole located at the core of the
M87 \cite{EHTL1,EHTL4} and SgrA*  \cite{EHT1,EHT3,EHT6} galaxies in 
accordance to predictions of general relativistic numerical simulations, supporting 
further the existence of these enigmatic black entities.

Within this black hole impulse we refine a general relativistic method that allows one to express the Kerr black hole parameters, mass and spin, as functions of directly observed quantities provided by astrometry and spectroscopy, namely, by the positions on the sky of particles revolving around the black hole in geodesic motion and the redshift experienced by the photons they emit when detected on Earth \cite{hanPRD2015}. This general relativistic formalism has been applied to several black hole metrics and compact objects in the literature so far. In \cite{Sharif}, the authors studied the frequency shifts of photons emitted  by particles near a Myers--Perry black hole with higher dimensions. A similar method was used in the case of the Kerr--Newman and Kerr--Newman--de Sitter black hole geometries \cite{Kraniotis} and the Plebanski--Demianski black hole \cite{Ujjal}.  This approach was used in \cite{Becerril1} to obtain the mass parameters of compact objects such as boson stars, as well as the Schwarzschild and Reissner-Nordström black holes from redshifts and blueshifts emitted by geodesic particles around them. In another work \cite{Becerril2}, the authors utilized a similar methodology to find the mass parameter of regular black holes and found the bounds on redshift and blueshifts of photons emitted by orbiting emitter particles. A generalization of this methodology was used to obtain the mass and the spin of a Kerr black hole in modified gravity \cite{Sheoran}, these authors used the redshift as a tool to test the Kerr black hole hypothesis. In \cite{Lopez1}, the authors calculated frequency shifts of photons emitted from geodesics of black holes with nonlinear electrodynamics, especially, the Bardeen and Bronnikov black holes and the Born--Infeld and Dymnikova black holes. Further, the study of the redshift of light emitted by particles orbiting a black hole immersed in a strong magnetic field was performed in \cite{Lopez2}. Redshift data could be essential to study the features of motion of objects such as individual stars and compact gas clouds as well as compact binary stars around black holes as discussed in \cite{Komarov,Dokuchaev}.

 All these attempts were based on the kinematic redshift, a fact that rendered involved formulas for the mass of those black holes. However, the kinematic redshift is not a directly measured observational quantity as the total redshift of photons is. In this work we make use of the total redshift expression in order to parameterize the mass and spin of the Kerr black hole, along with the orbital radius of the revolving body. In this way we obtain concise and elegant analytic relations that allow us to compute the mass and the spin parameters in terms of few observational data, for the cases when the black hole is static or moving with respect to a distant observer located on Earth. It is to be noted that \cite{Rashmi} has applied a similar approach in the case of the Kerr--Sen black hole without making use of the kinematic redshift in their calculations. In this paper, we show that the total redshift must also incorporate an additional contribution coming from the special relativistic boost generated by the relative motion of the Earth with respect to the black hole. We further elucidate the effect of the gravitational dragging produced by the black hole rotation on the light bending parameter and on the gravitational and the kinematic components of the total redshift, correcting previous expressions where this relevant effect was ignored. 
 
This novel approach allows us to extract a closed formula for the
gravitational redshift in a very clean and neat manner. In principle, this relation also
allows us to quantify this general relativistic effect for concrete real
astrophysical systems like the accretion disks with water masers orbiting
supermassive black holes of several active galactic nuclei if the precision
in the involved observations is high enough. We manage as well to
incorporate the special relativistic redshift associated with the motion of
a galactic black hole with respect to us. Thus, we are able to consistently
disentangle and quantify both the general and special relativistic redshifts
from the total frequency shift measured here on Earth. It is worthwhile to 
mention that, although these relativistic corrections have been considered 
in previous studies, they could not be identified and quantified properly. 
Hence, the contribution of the general and special relativistic corrections 
to the detected total redshift was obfuscated.

In addition, previous works dealt with redshifted or blueshifted photons which have the maximum light bending parameter. It occurs when the radial component of the 4-momentum of the detected photons is zero when they are emitted from the source. This also restricts the formalism to some specific source positions with respect to the line of sight. Naturally, a general formula is required that can incorporate the redshift or blueshift of photons emitted from sources in any arbitrary positions around a black hole. This paper also aims to develop a formalism incorporating a general light bending parameter corresponding to a photon emitted from an arbitrary source position.

The paper is organized as follows: We present a brief overview of our general relativistic 
method in Subsec. \ref{review}. We further derive concise analytic expressions
for the black hole mass and spin parameters as functions of directly
observable quantities in Sec. II.
In Sec. III, we take into account the dragging effect produced by the
rotating character of the Kerr black hole on the light bending parameter and
both the gravitational redshift and the kinematic frequency shift. In
Sec. IV, we boost the black hole with respect to a distant observer with the
aid of a composition of the Kerr redshift with a special relativistic
frequency shift. In Sec. V, we introduce the dependency of the Kerr redshift
and blueshift of photons on the azimuthal angle spanned by a probe particle,
allowing us to model the photons' frequency shift coming from a general point
in the equatorial plane. In Sec. VI, we also boost the Kerr black hole with
respect to a distant observer and compose the expressions for the frequency
shift with the incorporated azimuthal angle with a special relativistic
redshift that takes into account the black hole motion. Finally, in Sec.
VII, we conclude with some final remarks related to the application of the
developed general relativistic formalism to real astrophysical systems as
well as with a brief discussion of our results.

\subsection{Brief overview of the GR formalism}
\label{review}

Here we review previous results of our general relativistic method in order to 
place our original contributions in context.
We first consider the geodesic motion of massive probe particles orbiting a
Kerr black hole metric given by the following line element
\begin{equation}
ds^{2}=g_{tt}dt^{2}+2g_{t\varphi }dtd\varphi +g_{\varphi \varphi }d\varphi
^{2}+g_{rr}dr^{2}+g_{\theta \theta }d\theta ^{2}  \label{metric}
\end{equation}%
with the metric components
\begin{equation}
g_{tt}=-\left( 1-\frac{2Mr}{\Sigma }\right) ,\quad g_{t\varphi }=-\frac{%
2Mar\sin ^{2}\theta }{\Sigma },\quad g_{rr}=\frac{\Sigma }{\Delta },  \notag
\end{equation}%
\begin{equation}
g_{\varphi \varphi }=\left( r^{2}+a^{2}+\frac{2Ma^{2}r\sin ^{2}\theta }{%
\Sigma }\right) \sin ^{2}\theta \,,\quad g_{\theta \theta }=\Sigma \,,
\notag
\end{equation}%
where $M^{2}\geq a^{2}$, $\ \ g_{t\varphi }^{2}-g_{\varphi \varphi
}g_{tt}=\Delta \sin ^{2}\theta \ $ and
\begin{equation}
\Delta =r^{2}+a^{2}-2Mr\,,\quad \quad \Sigma =r^{2}+a^{2}\cos ^{2}\theta \,,
\notag
\end{equation}%
where $M$ is the total mass of the black hole and $a$ is the angular
momentum per unit mass, $a=J/M$ ($0\leq a\leq M$). The probe particles feel
the curvature of spacetime produced by the black hole through the metric and
keep memory of its parameters: the mass and the spin. On the other hand,
observers located on these particles can exchange electromagnetic signals
(photons) that travel along null geodesics from emission to detection and
have information of the aforementioned memory. Therefore, the frequency
shifts that these photons experience during their path, along with the
orbital parameters of the emitter and the observer can be used to determine
the mass and the spin parameters of the Kerr black hole according to the
inverse method introduced in \cite{hanPRD2015}. Thus, this formalism allows
one to compute the values of the Kerr black hole parameters on the basis of
directly measured observational quantities: the total redshifts and
blueshifts of the emitted photons and the positions of their source, in
contrast to the commonly used radial velocities, which are coordinate
dependent observables.

Within general relativity (GR), the frequency of a photon with 4-momentum $%
k_{c}^{\mu }=\left( k^{t},k^{r},k^{\theta },k^{\varphi }\right) \mid _{c}$,
which is emitted or detected by an emitter or an observer with proper 4-velocity $%
U_{c}^{\mu }=(U^{t},U^{r},U^{\theta },U^{\varphi })\mid _{c}$ at the point $%
c $, is a general relativistic invariant quantity that reads
\begin{equation}
\omega _{c}=-\left( k_{\mu }U^{\mu }\right) \mid _{c}\,,  \label{freq}
\end{equation}%
where the index $_{c}$ refers to the point of emission $_{e}$ or detection $%
_{d}$ of the photon. In the special case when the detector is located far
away from the emitter source, ideally at spatial infinity $(r\rightarrow
\infty )$, the 4-velocity simplifies
\begin{equation}
U_{d}^{\mu }=(1,0,0,0)\,.  \notag  \label{velocityd_infty}
\end{equation}

Besides, in axially symmetric backgrounds of the form (\ref{metric}), the
most general expression for the frequency shift that light signals emitted
by massive particles experience in their path along null geodesics towards a
detecting observer 
is given by the following relation \cite{hanPRD2015}
\begin{eqnarray}
1 &+&z_{_{Kerr}}\!=\frac{\omega _{e}}{\omega _{d}}  \notag \\
&=&\frac{(E_{\gamma }U^{t}-L_{\gamma }U^{\varphi
}-g_{rr}U^{r}k^{r}-g_{\theta \theta }U^{\theta }k^{\theta })\mid _{e}}{%
(E_{\gamma }U^{t}-L_{\gamma }U^{\varphi }-g_{rr}U^{r}k^{r}-g_{\theta \theta
}U^{\theta }k^{\theta })\mid _{d}}\,,
\end{eqnarray}%
where the conserved quantities $E_{\gamma }$ and $L_{\gamma }$ stand for the
total energy and axial angular momentum of the photon. This equation for the
redshifts and blueshifts includes stable orbits of any kind for the probe
particles (stars, for instance): Circular, elliptic, irregular, equatorial,
non-equatorial, etc. when moving around a Kerr black hole.


\subsubsection{Redshift of photons emitted by bodies in circular and equatorial orbits}

In order to explore the basic properties of accretion disks, studying the
equatorial circular motion of test particles in the background geometry of
the rotating black holes is inevitable because any tilted disk should be
driven to the equatorial plane of the rotating background \cite%
{BardeenPetterson}. Hence, we focus on the important case of circular and
equatorial orbits of probe massive particles, when $U^{r}=0=U^{\theta }$,
and present closed formulae for both the mass and rotation parameters of the
Kerr black hole in terms of measured redshifts and blueshifts of light
signals detected by an observer located far away from their source.

For the frequency shift of photons in this case, the general expression (3) adopts the 
form
\begin{equation}
1+z_{_{Kerr_{1,2}}}\!=\frac{\omega _{e}}{\omega _{d}}=\frac{\left. \left(
E_{\gamma }U^{t}-L_{\gamma }U^{\varphi }\right) \right\vert _{e}}{\left.
\left( E_{\gamma }U^{t}-L_{\gamma }U^{\varphi }\right) \right\vert _{d}}=%
\frac{U_{e}^{t}-b_{e_{(\mp )}}\,U_{e}^{\varphi }}{U_{d}^{t}-b_{d_{(\mp
)}}\,U_{d}^{\varphi }}\,,  \label{zcircorbits}
\end{equation}%
where the $_{(\mp )}$ subscripts denote two different values of the light
bending parameter that correspond to photons emitted by two different source positions 
either side of the line of sight; the subindices $_{_{1}}$ and $_{_{2}}$ correspond to the $_{_{(-)}}$
and $_{_{(+)}}$ signs, respectively. Besides, the deflection of light
parameter $b$ is defined by $b\equiv L_{\gamma }/E_{\gamma }$ and it takes
into account the light bending generated by the gravitational field in the
vicinity of the rotating black hole. This parameter is preserved along the
whole null geodesics followed by photons from their emission till their
detection, and we have $b_{e}=b_{d}$ since both $E_{\gamma }$ and $L_{\gamma
}$ are constants of motion.

In a natural form, one sees that $z_{_{Kerr_{1}}}\neq z_{_{Kerr_{2}}}$
by definition. In fact, this difference has two reasons: ($i$) the
gravitational redshift produced by the black hole mass and its angular
momentum, which is always positive, and ($ii$)
different light bending parameters $b_{e_{(\mp )}}$ experienced by the
emitted photons on either side of the line of sight (in both cases, when the 
photon source is co-rotating and counter-rotating with respect to 
the angular velocity of the black hole). Thus, the gravitational
field bends the light in a different way for approaching and receding photon
sources due to these general relativistic effects.

\begin{figure*}[tbp]
	\centering
	\includegraphics[width=0.5\textwidth]{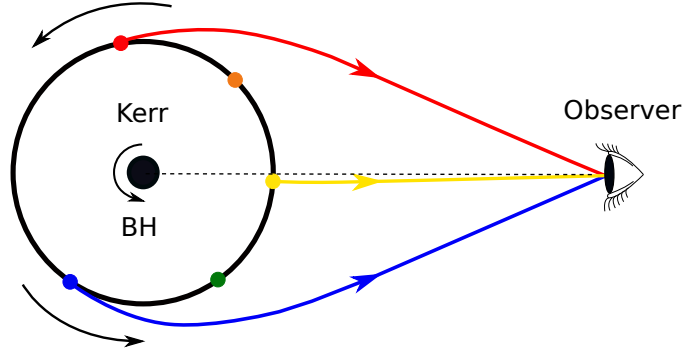}
	\caption{Schematic diagram depicts the frequency shifts of three different light rays emitted by timelike sources orbiting a Kerr black hole in an equatorial, circular geodesic and detected by an observer also located on the equatorial plane. The blue and red trajectories are emitted with $ k^r = 0 $ indicating maximum frequency shifts, whereas the yellow ray is radially emitted, i.e.,  $k^{\varphi} = 0 $ at its emission point. Due to the frame dragging of the Kerr background, trajectories are asymmetric with respect to the line of sight (dashed line). The colored points indicate different amount of frequency shifts depending on the different positions of the sources.  }
	\label{Schematic}
\end{figure*}

The maximum value of the light bending parameter is attained when $k^{r}=0$,
where the position vectors of orbiting objects with respect to the black hole
location are approximately orthogonal to the line of sight (see Fig. \ref{Schematic}), rendering for the
Kerr metric \cite{hanPRD2015}
\begin{eqnarray}
b_{(\pm )} &=&-\frac{g_{t\varphi }(\pm )\sqrt{g_{t\varphi
}^{2}-g_{tt}g_{\varphi \varphi }}}{g_{tt}}  \notag \\
&=&\frac{-2Ma(\pm )r\sqrt{r^{2}-2Mr+a^{2}}}{r-2M},  \label{LBParameter}
\end{eqnarray}%
where the latter equality stands for photons emitted by particles in circular equatorial orbits. It is
notable to mention that, as we shall see, the sign of $b$ characterizes
whether a photon is redshifted or blueshifted when the particle is 
co-rotating with respect to the black hole (and viceversa when it is 
counter-rotating). Therefore, from now on, the
minus sign enclosed in parentheses denotes the redshifted photons, whereas
the plus sign represents blueshifted ones in the frequency shift formulas.

On the other hand, the non-vanishing components of the $4$-velocity for
circular equatorial orbits read
\begin{equation}
U^{t}(r,\pi /2)=\frac{r^{3/2}\pm aM^{1/2}}{\mathcal{X}_{\pm }},
\label{tCompVelo}
\end{equation}%
\begin{equation}
U^{\varphi }(r,\pi /2)=\frac{\pm M^{1/2}}{\mathcal{X}_{\pm }},
\label{phiCompVelo}
\end{equation}%
with $\mathcal{X}_{\pm }=r^{3/4}\sqrt{r^{3/2}-3Mr^{1/2}\pm 2aM^{1/2}}$. In
these relations, the upper sign corresponds to a co-rotating object while
the lower sign refers to a counter-rotating one, and we use this convention
in the upcoming equations. By considering Eqs. (\ref{tCompVelo}) and (\ref%
{phiCompVelo}), we can also obtain the angular velocity of an object
orbiting around the Kerr black hole as below%
\begin{equation}
\Omega _{\pm }=\frac{d\varphi }{dt}=\frac{U^{\varphi }}{U^{t}}=\frac{\pm
M^{1/2}}{r^{3/2}\pm aM^{1/2}},  \label{AngularVelocity}
\end{equation}%
which acquires an additional subscript $e/d$ for the emitter/detector as
well.

\section{Black hole mass and spin from redshift/blueshift of photons in
equatorial orbits}

With the quantities presented in the previous Section at hand, we now 
express the frequency shift $z_{_{Kerr_{1,2}}}$ and obtain the mass and 
angular momentum parameters of the Kerr black hole in terms of the 
redshift and blueshift of the photons emitted by orbiting objects and their 
orbital radii. By substituting (\ref{LBParameter})-(\ref{phiCompVelo}) into 
(\ref{zcircorbits}), we obtain
\begin{equation}
1+z_{_{Kerr_{1}}}\!=\frac{\mathcal{X}_{d\pm }\left[ r_{e}^{3/2}\left(
r_{e}-2M\right) \pm M^{1/2}r_{e}\left( a+\sqrt{\Delta _{e}}\right) \right] }{%
\mathcal{X}_{e\pm }\left[ r_{d}^{3/2}\left( r_{e}-2M\right) \pm
M^{1/2}r_{e}\left( a+\sqrt{\Delta _{e}}\right) \right] },
\end{equation}%
\begin{equation}
1+z_{_{Kerr_{2}}}\!=\frac{\mathcal{X}_{d\pm }\left[ r_{e}^{3/2}\left(
r_{e}-2M\right) \pm M^{1/2}r_{e}\left( a-\sqrt{\Delta _{e}}\right) \right] }{%
\mathcal{X}_{e\pm }\left[ r_{d}^{3/2}\left( r_{e}-2M\right) \pm
M^{1/2}r_{e}\left( a-\sqrt{\Delta _{e}}\right) \right] },
\end{equation}%
where $\mathcal{X}_{c\pm }=\left. \mathcal{X}_{\pm }\right\vert _{r=r_{c}}$,
$\Delta _{c}=\left. \Delta \right\vert _{r=r_{c}}$, $r_{e}$ ($r_{d}$)
denotes the radius of the emitter (detector) orbit, and we used $b_{d}=b_{e}$%
. In the particular case, when the detector is located far away from the
source ($r_{d}>>M\geq a$ and $r_{d}>>r_{e}$), these relations reduce to
\begin{equation}
1+z_{_{Kerr_{1}}}\!=\frac{\left( 1-2\tilde{M}\right) \pm
\tilde{M}^{1/2}\left( \tilde{a}+\sqrt{\tilde{\Delta}_{e}}\right) }
{\left(1-2\tilde{M}\right) \sqrt{1-3\tilde{M}\pm 2\,\tilde{a}\,\tilde{M}^{1/2}}},
\label{KerrRedFarDetector}
\end{equation}%
\begin{equation}
1+z_{_{Kerr_{2}}}\!=\frac{\left( 1-2\tilde{M}\right) \pm
\tilde{M}^{1/2}\left( \tilde{a}-\sqrt{\tilde{\Delta}_{e}}\right) }
{\left(1-2\tilde{M}\right) \sqrt{1-3\tilde{M}\pm 2\,\tilde{a}\,\tilde{M}^{1/2}}},
\label{KerrBlueFarDetector}
\end{equation}
with $\tilde{M}=M/r_{e}$, $\ \tilde{a}=a/r_{e}$ and $\tilde{\Delta}_{e}=1+\tilde{a}^{2}-2\tilde{M}$.

Now, it is straightforward to show that%
\begin{equation}
RB=\frac{1}{1-2\tilde{M}},  \label{M}
\end{equation}%
\begin{equation}
\frac{R}{B}=\frac{1-2\tilde{M}\pm\tilde{M}^{1/2}\tilde{a}\pm\tilde{M}^{1/2}%
\sqrt{1+\tilde{a}^{2}-2\tilde{M}}}{1-2\tilde{M}\pm\tilde{M}^{1/2}\tilde{a}\mp%
\tilde{M}^{1/2}\sqrt{1+\tilde{a}^{2}-2\tilde{M}}},  \label{RoverB}
\end{equation}
where we introduced $R=1+z_{_{Kerr_1}}$ and $B=1+z_{_{Kerr_2}}$.

In this case, we are able to find a closed formula for the mass of the black
hole in terms of observational quantities, namely, the total redshift and
blueshift of photons emitted by particles revolving the black hole as well
as their orbital radius
\begin{equation}
M=\frac{RB-1}{2RB}r_{e}=\frac{(1+z_{_{Kerr_{1}}})(1+z_{_{Kerr_{2}}})-1}{%
2(1+z_{_{Kerr_{1}}})(1+z_{_{Kerr_{2}}})}r_{e}.  \label{mass}
\end{equation}

Since the right-hand side of the derivation of the black hole mass in (\ref{M}) 
does not involve the angular momentum parameter, it is valid as well for the 
Schwarzschild black hole mass and it can be generalized straightforwardly 
for any spherically symmetric metric depending on more free parameters.

We would like to emphasize that the mass expression (\ref{mass}) is given in
terms of the total Kerr frequency shifts (\ref{KerrRedFarDetector})-(\ref%
{KerrBlueFarDetector}) which are directly measured quantities in real
astrophysical systems. This elegant relation differs from previous
attempts to express the mass parameter in terms of the kinematic frequency
shift \cite{hanPRD2015,Becerril1,Becerril2,halcnAN2021}. Indeed, since Eq. (\ref{M}) is a
linear relation for $M$, it has no degeneracy in its values as in the case
of higher order algebraic equations arising in \cite{Becerril1,Becerril2} when expressing $M$ in terms of the kinematic redshift for several spherically symmetric spacetimes and for configurations involving
several photon sources \cite{halcnAN2021}.

Now, by substituting the relation (\ref{mass}) into (\ref{RoverB}), we
obtain the rotation parameter versus the same quantities%
\begin{equation}
|a|=\frac{\left( R-B\right) ^{2}-\left( R+B\right) \sqrt{%
R^{2}+B^{2}-2R^{2}B^{2}}}{\left( 2RB\right) ^{3/2}\sqrt{RB-1}}r_{e}.
\label{AngularAB}
\end{equation}

It is also straightforward to show that the expression for the rotation
parameter is consistent with the bound $M\ge|a|$.

By computing the ratio $|a|/M$, we get rid of the orbital radius dependence
and obtain a bounded expression for the spin parameter per unit mass in
terms of observational frequency shifts
\begin{equation}
\frac{|a|}{M}=\frac{\left( R-B\right) ^{2}-\left( R+B\right) \sqrt{%
R^{2}+B^{2}-2R^{2}B^{2}}}{\sqrt{2RB}\ \left( RB-1\right) ^{3/2}},
\end{equation}%
where $0\leq |a|/M\leq 1$.

We should note that there are some constraints on $R$\ and $B$\ to have
positive definite $M$ and $a$ in the relations (\ref{mass})\ and (\ref%
{AngularAB}), respectively. These conditions are $RB>1$\ and $%
R^{2}+B^{2}\geq 2R^{2}B^{2}$\ which must be obeyed. On the other hand,\ the
radius of innermost stable circular orbit (ISCO) in the Kerr geometry is
given by \cite{Bardeen}
\begin{equation}
r_{ms}=M\left( 3+\beta \mp \sqrt{(3-\alpha )(3+\alpha +2\beta )}\right) ,
\label{rms}
\end{equation}%
\begin{equation*}
\alpha =1+\left( 1-\frac{a^{2}}{M^{2}}\right) ^{1/3}\left[ \left( 1+\frac{a}{%
M}\right) ^{1/3}+\left( 1-\frac{a}{M}\right) ^{1/3}\right] ,
\end{equation*}%
\begin{equation*}
\beta =\sqrt{\alpha ^{2}+3\frac{a^{2}}{M^{2}}},
\end{equation*}%
that approximately characterizes the inner edge of orbiting accretion disk
and \textquotedblleft $ms$\textquotedblright\ stands for \textquotedblleft
marginally stable\textquotedblright\ orbit. Thus, the lower bound on the
emitter radius as $r_{e}\geq r_{ms}$\ leads to a bound on the
redshift/blueshift presented in (\ref{KerrRedFarDetector})\ and (\ref%
{KerrBlueFarDetector}).

\begin{figure*}[tbp]
\centering
\includegraphics[width=0.4\textwidth]{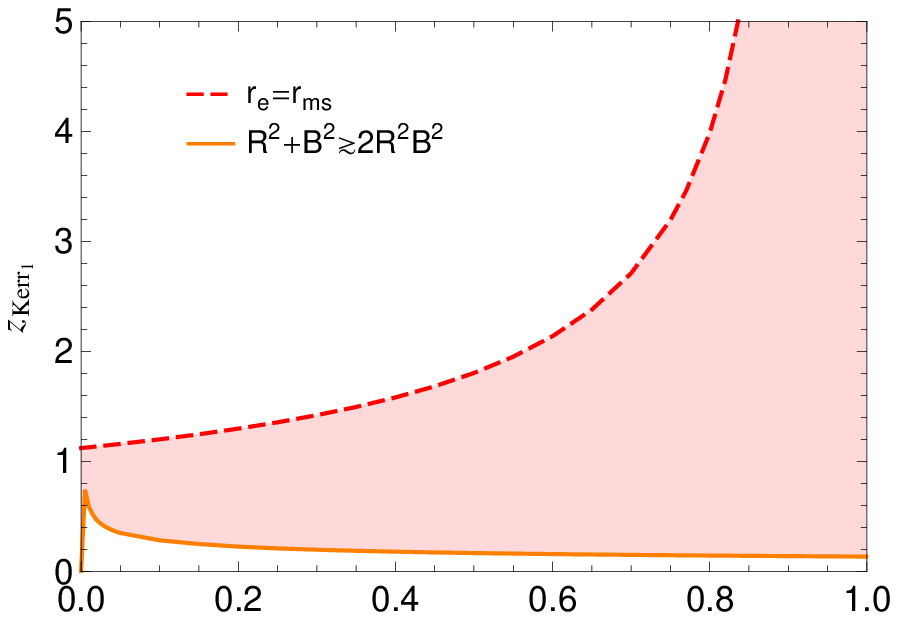} %
\includegraphics[width=0.4\textwidth]{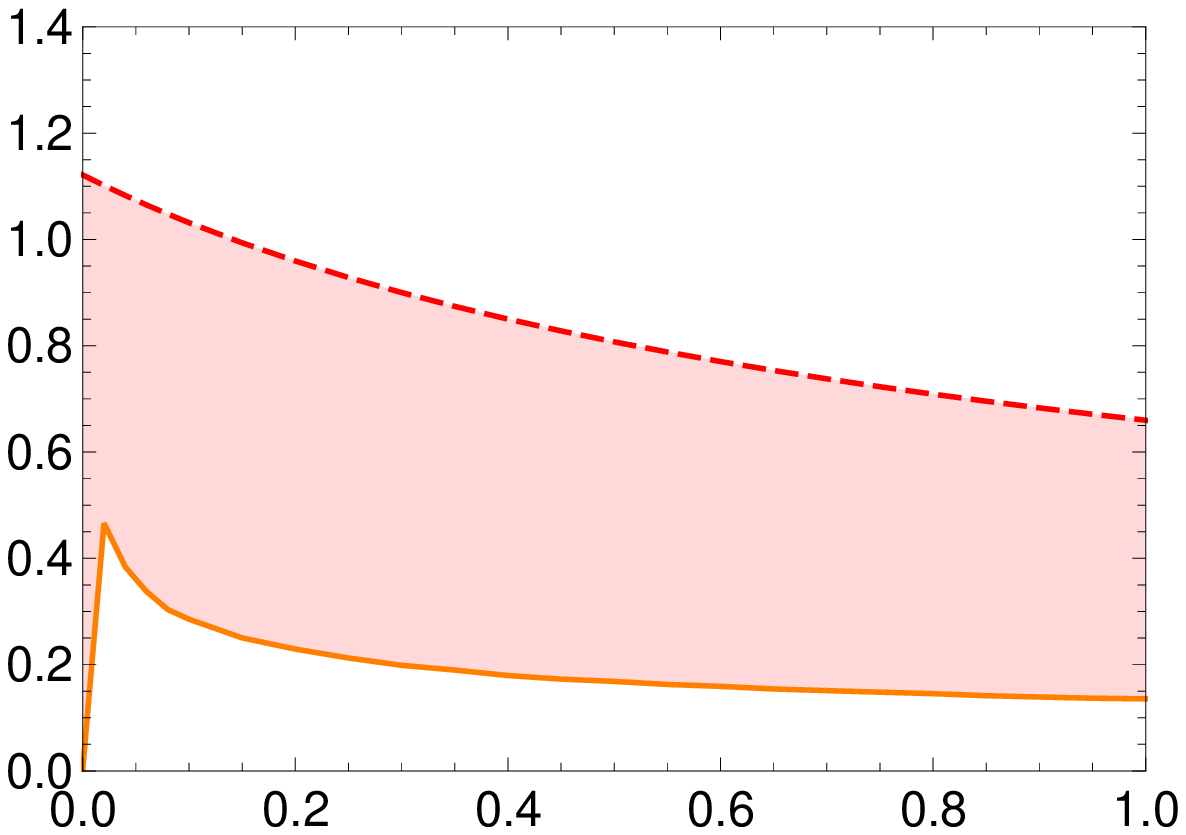} %
\includegraphics[width=0.41\textwidth]{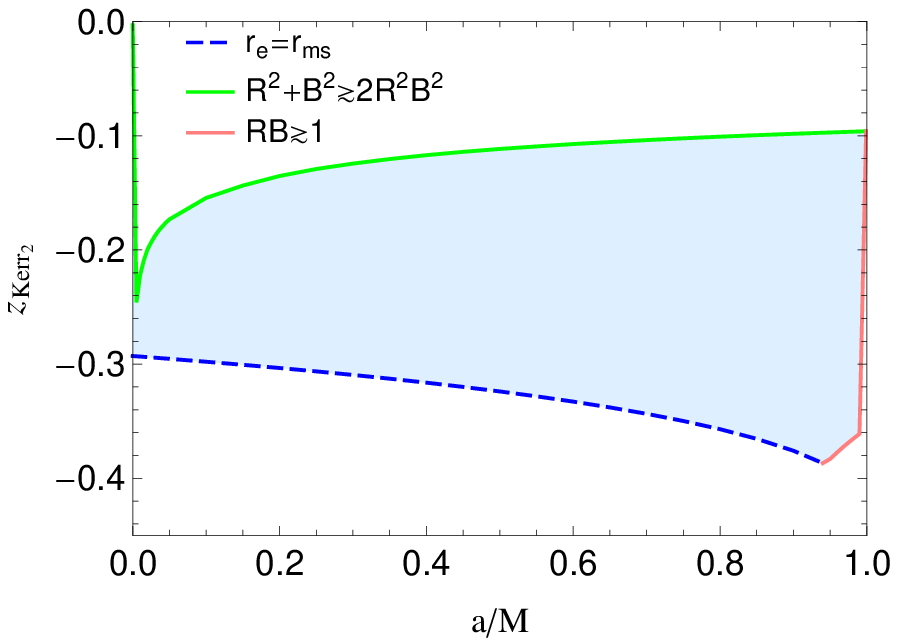} %
\includegraphics[width=0.41\textwidth]{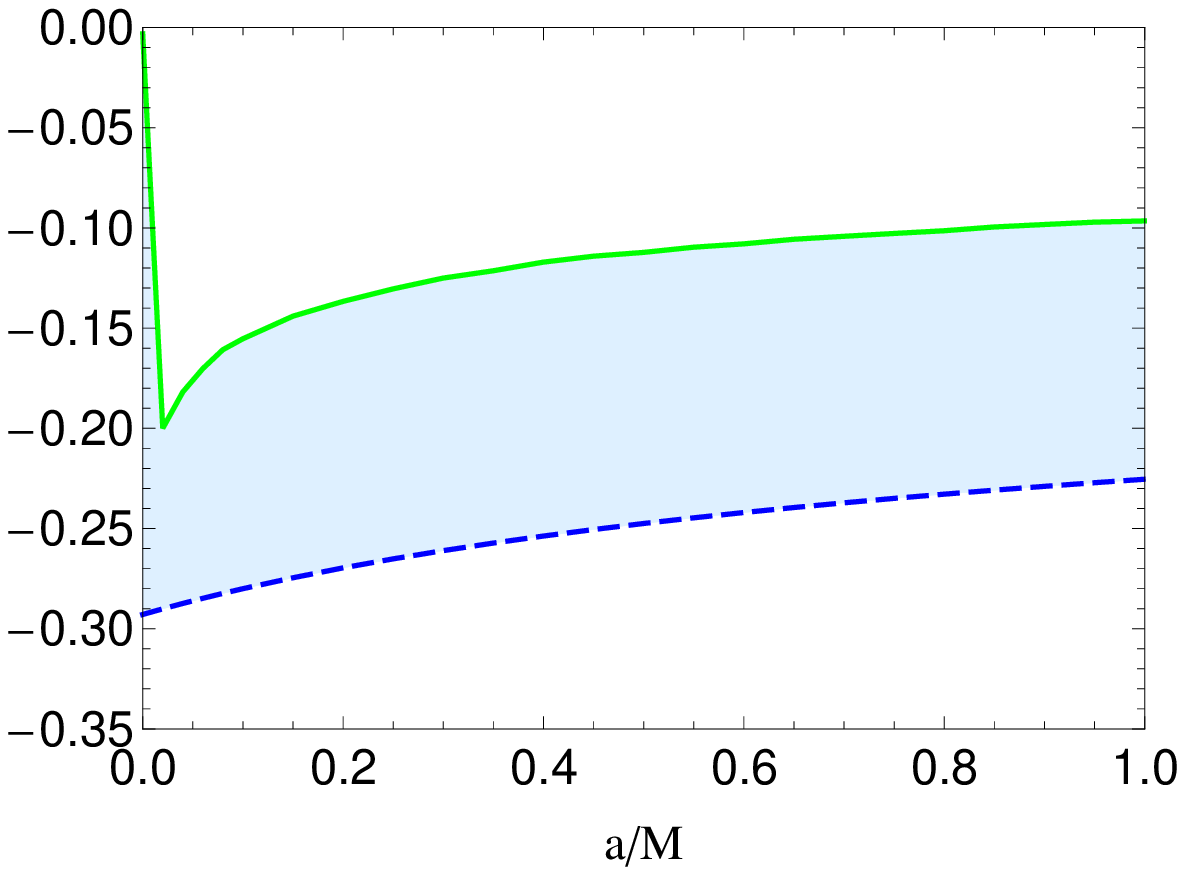}
\caption{The redshift $z_{_{Kerr_{1}}}$ and blueshift $z_{_{Kerr_{2}}}$
versus the rotation parameter for co-rotating branch (left panels) and
counter-rotating branch (right panels). The dashed curves (denoted by $%
r_{ms} $) correspond to the value of redshift/blueshift emitted by geodesic
particles on marginally stable orbit $r_{ms}$ (ISCO). The continuous curves
show the redshift/blueshift values as $R^{2}+B^{2}\rightarrow 2R^{2}B^{2}$
and $RB\rightarrow 1$ from above. The shaded region indicates the valid
values of $z_{_{Kerr_{1}}}$ and $z_{_{Kerr_{2}}}$ for the Kerr black hole.}
\label{KerrShiftFig}
\end{figure*}

The redshift (\ref{KerrRedFarDetector})\ and blueshift (\ref%
{KerrBlueFarDetector}) of photons emitted by geodesic particles orbiting the
Kerr black hole are illustrated in Fig. \ref{KerrShiftFig}. This figure
shows that the redshift is bounded from the top by marginally stable orbits
and from the bottom by the condition $R^{2}+B^{2}\geq 2R^{2}B^{2}$, and
also, $RB>1$ in the case of co-rotating objects. However, the blueshift is
bounded from the bottom by marginally stable orbits and from the top by the
condition $R^{2}+B^{2}\geq 2R^{2}B^{2}$. Therefore, generally, the shaded
area indicates the valid values for the redshift and blueshift of photons
radiated by stable objects orbiting the Kerr black hole, bounded between two
curves characterized by $r_{ms}$\ and the condition $R^{2}+B^{2}\geq
2R^{2}B^{2}$ (and also, $RB>1$ in the case of co-rotating objects).

Note that, for the co-rotating branch, the absolute value of the
redshift/blueshift at the ISCO (dashed curves in the left panels of Fig. \ref%
{KerrShiftFig}) is an increasing function of $a$, whereas for the
counter-rotating case, $\left\vert z_{_{Kerr_{1,2}}}\right\vert $ is a
decreasing function of $a$ (see the dashed curves in the right panels of
Fig. \ref{KerrShiftFig}). The dependency of $z_{_{Kerr_{1,2}}}$ on the
rotation parameter is a consequence of the dragging effect produced by the
rotation nature of the Kerr black hole that we shall investigate in detail.

\section{The dragging effect on the light bending parameter and the redshift}

In contrast to the Schwarzschild black hole case, a probe particle cannot be
static in the Kerr background due to the dragging effect generated by rotation.
Thus, a massive particle located on the line of sight in its circular orbit
necessarily feels a drag that modifies the corresponding redshift experienced 
by the emitted photons when detected by a distant observer.

Therefore, we now develop our formalism to take into account the dragging
effect produced by the rotating nature of the Kerr metric on the central
light bending parameter, the gravitational redshift $z_{g}$, and kinematic
redshifts and blueshifts $z_{kin_{\pm }}$. The latter quantities encode the
rotational motion of probe particles around the black hole. The inclusion of
this effect was missing in the analysis performed in \cite{hanPRD2015} and
all subsequent works which considered rotating metrics (see \cite%
{Sheoran,Kraniotis
}, for instance). Here we fill this important gap.

It is worthwhile to mention that the light bending parameter $ b_c $ of a light ray emitted radially at the central point (on the line of sight) is non-zero due to the dragging effect. Therefore, a light ray which is emitted radially at the central point, bends due to the dragging effect and cannot reach the observer located on the line of sight. This also means that the photon must be emitted from a point away from the central point so that it can reach the observer and deliver to it information about the gravitational redshift (See Fig. 1).  This quantity has no analog in the Newtonian picture and we
are going to take this effect into account through this section.

In general, for a stationary axisymmetric spacetime, the following
components of the $4$-momentum of photons can be expressed in terms of the
metric components and the constants of motion as follows \cite{Bardeen,Wilkins}
\begin{equation}
k^{t}=\frac{E_{\gamma }g_{\varphi \varphi }+L_{\gamma }g_{t\varphi }}{%
g_{t\varphi }^{2}-g_{tt}g_{\varphi \varphi }},  \label{tCompOf4momentum}
\end{equation}%
\begin{equation}
k^{\varphi }=-\frac{E_{\gamma }g_{t\varphi }+L_{\gamma }g_{tt}}{g_{t\varphi
}^{2}-g_{tt}g_{\varphi \varphi }}.  \label{PhiCompOf4momentum}
\end{equation}%
The $\varphi $-component vanishes for a radially emitted photon, a fact
that leads to the following relation for the light bending parameter%
\begin{equation}
b_{c}=-\frac{g_{t\varphi }}{g_{tt}}=-\frac{2Ma}{r-2M},  \label{DragBendLigh}
\end{equation}%
where the second equality takes place for photons emitted by bodies in 
circular and equatorial orbits
around the Kerr black hole. Here, the effect of the dragging due to the
rotating character of the Kerr black hole spacetime becomes clear since 
$b_{c}$ is proportional to $a$ and it vanishes for a Schwarzschild black 
hole configuration.

Thus, by considering the light bending parameter of a radially emitted photon, the
gravitational redshift of circular motion in the equatorial plane around the
Kerr black hole becomes%
\begin{equation}
1+z_{g}=\frac{U_{e}^{t}-b_{c}U_{e}^{\varphi }}{U_{d}^{t}-b_{c}U_{d}^{\varphi
}}=\frac{1-2\tilde{M}\pm \tilde{M}^{1/2}\tilde{a}}{\left( 1-2\tilde{M}%
\right) \sqrt{1-3\tilde{M}\pm 2\tilde{M}^{1/2}\tilde{a}}},
\label{CentralRedshift}
\end{equation}%
where the last part is obtained for a far away detector ($r_{d}>>M\geq a$
and $r_{d}>>r_{e}$). This expression depends on both the mass and the spin
parameters, implying the fact that the gravitational redshift is generated by the
black hole mass (given by the first two items when the spin vanishes) and
its rotation (the third term). This quantity generalizes the gravitational
redshift formula for the Schwarzschild black hole given in
\cite{nhalclcAPJL2021} and reproduces it when the rotation parameter is trivial.

On the other hand,
the kinematic redshift $z_{_{kin_{\pm}}}$ is defined by subtracting from the
Kerr redshift $z_{_{Kerr_{1,2}}}$, the gravitational redshift $z_{g}$
experienced by photons radially emitted,
such that $b=b_{c} $, as follows%
\begin{equation}
z_{_{kin_{\pm}}}=z_{_{Kerr_{1,2}}}\!-z_{g}= \frac{U^t_e - b_{e_{(\mp)}}
\,U^\varphi_e}{U^t_d - b_{d_{(\mp)}} \,U^\varphi_d}-\frac{%
U_{e}^{t}-b_{c}U_{e}^{\varphi }}{U_{d}^{t}-b_{c}U_{d}^{\varphi }}.
\label{KinZ}
\end{equation}

We recall that the upper/lower sign in $z_{_{Kerr_{1,2}}}$ (or $U^{t}$ and $%
U^{\varphi }$) corresponds to a co/counter-rotating object while the
minus/plus sign enclosed in parentheses denotes the redshifted/blueshifted
photons. Thus, by taking into account the maximum values of the light
bending parameter (\ref{LBParameter}) on either side of the line of sight,
we are led to two different kinematic frequency shifts for receding and
approaching objects.%

By using the definition of the angular velocity (\ref{AngularVelocity}) of
the detector ($\Omega _{d}=U_{d}^{\varphi }/U_{d}^{t}$) and emitter ($\Omega
_{e}=U_{e}^{\varphi }/U_{e}^{t}$), the kinematic frequency shifts $%
z_{_{kin_{+}}}$\ and $z{_{kin_{-}}}$\ take the following form%
\begin{equation}
z_{_{kin_{+}}}=\frac{U_{e}^{t}}{U_{d}^{t}}\left( \frac{1-b_{e_{-}}\Omega _{e}%
}{1-b_{d_{-}}\Omega _{d}}-\frac{1-b_{c}\Omega _{e}}{1-b_{c}\Omega _{d}}%
\right) ,  \label{KinZ1}
\end{equation}%
\begin{equation}
z_{_{kin_{-}}}=\frac{U_{e}^{t}}{U_{d}^{t}}\left( \frac{1-b_{e_{+}}\Omega _{e}%
}{1-b_{d_{+}}\Omega _{d}}-\frac{1-b_{c}\Omega _{e}}{1-b_{c}\Omega _{d}}%
\right) .  \label{KinZ2}
\end{equation}

Alternatively, by substituting the $4$-velocity components (\ref{tCompVelo}%
)\ and (\ref{phiCompVelo})\ into these kinematic redshifts $z_{_{kin_{\pm }}}$
(Eqs. (\ref{KinZ1}) and (\ref{KinZ2})), we obtain the following
expressions 
\begin{equation}
\label{KinZ+}
z_{_{kin_{+}}}=\frac{\mathcal{X}_{d\pm }\Omega _{d\pm }}{\mathcal{X}_{e\pm
}\Omega _{e\pm }}\left( \frac{1-b_{e_{-}}\Omega _{e\pm }}{1-b_{d_{-}}\Omega
_{d\pm }}-\frac{1-b_{c}\Omega _{e\pm }}{1-b_{c}\Omega _{d\pm }}\right) ,
\end{equation}%
\begin{equation}
\label{KinZ-}
z_{_{kin_{-}}}=\frac{\mathcal{X}_{d\pm }\Omega _{d\pm }}{\mathcal{X}_{e\pm
}\Omega _{e\pm }}\left( \frac{1-b_{e_{+}}\Omega _{e\pm }}{1-b_{d_{+}}\Omega
_{d\pm }}-\frac{1-b_{c}\Omega _{e\pm }}{1-b_{c}\Omega _{d\pm }}\right) .
\end{equation}

By directly substituting (\ref{LBParameter})-(\ref{phiCompVelo}) and (\ref%
{DragBendLigh}) into (\ref{KinZ}) [or, equivalently, by employing (\ref%
{LBParameter}), (\ref{AngularVelocity}), and (\ref{DragBendLigh}) in relations 
(\ref{KinZ+})-(\ref{KinZ-})], it is possible to write these expressions 
{\it versus} the black hole parameters and orbital radii%
\begin{eqnarray}
&&z_{_{kin_{+}}}= \\
&&\frac{\pm M^{\frac{1}{2}}\mathcal{X}_{d\pm }\left( r_{d}^{\frac{3}{2}%
}-r_{e}^{\frac{3}{2}}\right) \left( r_{e}\!-\!2M\right) \sqrt{\Delta _{e}}r_{e}}{%
\mathcal{X}_{e\pm }\!\!\left[ r_{d}^{\frac{3}{2}}\!\left(
r_{e}\!\!-\!\!2M\right) \!\pm \!aM^{\frac{1}{2}}r_{e}\right] \!\!\left[
r_{d}^{\frac{3}{2}}\left( r_{e}\!\!-\!\!2M\right) \!\pm \!M^{\frac{1}{2}%
}r_{e}\!\left( a\!+\!\Delta _{e}\right) \right] },  \notag  \label{KinRed}
\end{eqnarray}%
\begin{eqnarray}
&&z_{_{kin_{-}}}= \\
&&\frac{\mp M^{\frac{1}{2}}\mathcal{X}_{d\pm }\left( r_{d}^{\frac{3}{2}%
}-r_{e}^{\frac{3}{2}}\right) \left( r_{e}\!-\!2M\right)\sqrt{\Delta _{e}}r_{e}}{%
\mathcal{X}_{e\pm }\!\!\left[ r_{d}^{\frac{3}{2}}\!\left(
r_{e}\!\!-\!\!2M\right) \!\pm \!aM^{\frac{1}{2}}r_{e}\right] \!\!\left[
r_{d}^{\frac{3}{2}}\left( r_{e}\!\!-\!\!2M\right) \!\pm \!M^{\frac{1}{2}%
}r_{e}\!\left( a\!+\!\Delta _{e}\right) \right] },  \notag  \label{KinBlue}
\end{eqnarray}%
where we used $b_{d}=b_{e}$. For a far away detector ($r_{d}>>M\geq a$ and $%
r_{d}>>r_{e}$), the kinematic redshift and blueshift take the following
simple forms%
\begin{eqnarray}
\label{KinRedFarDetector}
z_{_{kin_{+}}} 
&=&\frac{\pm \tilde{M}^{1/2}\sqrt{1+\tilde{a}^{2}-2\tilde{M}}}{\left(
1-2\tilde{M}\right) \sqrt{1-3\tilde{M}\pm 2\,\tilde{a}\,\tilde{M}^{1/2}}}, 
\end{eqnarray}%
\begin{eqnarray}
\label{KinBlueFarDetector}
z_{_{kin_{-}}} 
&=&\frac{\mp \tilde{M}^{1/2}\sqrt{1+\tilde{a}^{2}-2\tilde{M}}}{\left(
1-2\tilde{M}\right) \sqrt{1-3\tilde{M}\pm 2\,\tilde{a}\,\tilde{M}^{1/2}}}, 
\end{eqnarray}%
that are symmetric with respect to the line of sight such that $%
z_{_{kin_{+}}}=-z_{_{kin_{-}}}$ as one would expect. 
These symmetric
expressions contrast with the asymmetric ones previously obtained in \cite%
{hanPRD2015} when the gravitational dragging effect due to the spin of the
Kerr black hole metric was ignored.

\section{Boosting the black hole with respect to a distant observer}

\label{boosting}

At this stage, we further compose the Kerr shift $z_{_{Kerr_{1,2}}}%
$\ (or equivalently Eqs. (\ref{KerrRedFarDetector}) and (\ref%
{KerrBlueFarDetector})) with the redshift describing the relative motion
\footnote{%
The relative motion of a galaxy with respect to us is described by its
peculiar velocity if it is not within the Hubble flow. This peculiar
velocity is due to a local gravitational attraction generated by a galaxy
cluster, for instance, that draws the galaxy either toward us or away from us.} of a black
hole from a distant observer, $z_{boost}$, which is associated with a
special relativistic boost
\begin{equation}
1+z_{boost}=\gamma \left( 1+\beta \right) ;\ \ \gamma =\left( 1-\beta
^{2}\right) ^{-1/2},\ \ \beta =\frac{v_{0}}{c},
\end{equation}%
where $v_{0}=z_{0}c$ is the radial peculiar velocity of the black hole with respect to a far away observer. %
Here we are neglecting a possible transversal component in the black hole
motion which would be taken into account by the relation 
\begin{equation}
1+z_{boost}=\gamma \left[ 1+\cos\left(\delta \right)\beta \right] ,  \notag
\end{equation}%
where $\delta$ is an angle that codifies the special relativistic transverse Doppler shift \cite{Jackson}, introducing one more parameter in our model.
For the sake of simplicity we shall consider just the galactic motion projected along the line of
sight, setting $\delta$ to zero.

We shall call $z_{0}$ the peculiar redshift since it encodes the receding
from us or approaching toward us motion of the black hole. Astronomers
usually call this motion {\it systemic} when referring to orbiting the black
hole particles lying close to the line of sight, where the radial component
of the rotational velocity vanishes.

Now, by considering the peculiar redshift, the total redshift becomes \cite%
{nhalclcAPJL2021}%
\begin{equation}
z_{_{tot_{1,2}}}=\left( 1+z_{_{kin_{\pm }}}+z_{g}\right) \left(
1+z_{boost}\right) -1,  \label{boost}
\end{equation}%
in which the kinematic shifts $z_{_{kin_{\pm }}}$ are given in (\ref%
{KinRedFarDetector}) and (\ref{KinBlueFarDetector}) respectively, and the
gravitational redshift $z_{g}$\ is taken from (\ref{CentralRedshift}). By
substituting the expressions of $z_{_{kin_{\pm }}}$\ and $z_{g}$\ for
different branches including the co-rotating/counter-rotating ($+/-$) bodies
in the equations mentioned above, we find their following explicit form%
\begin{eqnarray}
&&z_{_{tot_{1}}}=  \label{TotRed} \\
&&\frac{\left( 1-2\tilde{M}\right) \pm \tilde{M}^{1/2}\left( \tilde{a}+\sqrt{%
1+\tilde{a}^{2}-2\tilde{M}}\right) }{\left( 1-2\tilde{M}\right) \sqrt{1\pm 2%
\tilde{a}\tilde{M}^{1/2}-3\tilde{M}}}\sqrt{\frac{1+z_{0}}{1-z_{0}}}-1,
\notag
\end{eqnarray}%
\begin{eqnarray}
&&z_{_{tot_{2}}}=  \label{TotBlue} \\
&&\frac{\left( 1-2\tilde{M}\right) \pm \tilde{M}^{1/2}\left( \tilde{a}-\sqrt{%
1+\tilde{a}^{2}-2\tilde{M}}\right) }{\left( 1-2\tilde{M}\right) \sqrt{1\pm 2%
\tilde{a}\tilde{M}^{1/2}-3\tilde{M}}}\sqrt{\frac{1+z_{0}}{1-z_{0}}}-1.
\notag
\end{eqnarray}

From these relations, the general effect of the peculiar redshift $z_{0}$ on
the Kerr shift $z_{_{Kerr_{1,2}}}$ is obvious. For $z_{0}>0$, when the black
hole is receding from us, we have $z_{_{tot_{1}}}>z_{_{Kerr_{1}}}$\ for
redshift and $\left\vert z_{_{tot_{2}}}\right\vert <\left\vert
z_{_{Kerr_{2}}}\right\vert $ for blueshift. But, for $z_{0}<0$, when the
black hole is approaching toward us, we have $z_{_{tot_{1}}}<z_{_{Kerr_{1}}}$\
for redshift and $\left\vert z_{_{tot_{2}}}\right\vert >\left\vert
z_{_{Kerr_{2}}}\right\vert $ for blueshift. Therefore, in the case of the
total shifts $z_{_{tot_{1,2}}}$, the shaded region of Fig. \ref{KerrShiftFig}
moves upward (downward) for $z_{0}>0$ ($z_{0}<0$) while the borders are
specified by (\ref{rms}), $\hat{R}\hat{B}\left( 1-z_{0}\right) ^{2}>\left(
1-z_{0}^{2}\right) $, and $G^{2}\geq 0$ (we shall introduce $\hat{R}$, $\hat{%
B}$, and $G$ below). Besides, note that $z_{_{tot_{1}}}$\ and $z_{_{tot_{2}}}$\
reduce to the corresponding relations (\ref{KerrRedFarDetector}) and (\ref%
{KerrBlueFarDetector}), respectively, when the peculiar redshift $z_{0}$ vanishes.

The total frequency shifts (\ref{TotRed})-(\ref{TotBlue}) for 
large orbits of the emitter $\tilde{M},\tilde{a}%
<<1 $\ read%
\begin{eqnarray}
&&
z_{_{tot_{1}}}\approx 
-1+\sqrt{\frac{1+z_{0}}{1-z_{0}}}\left( 1 \pm \sqrt{\tilde{M}}
+\right.  \notag \\
&&\left. \frac{3}{2}\tilde{M} \pm \frac{5}{2}\tilde{M}^{3/2} +\frac{27%
}{8}\tilde{M}^{2} - \tilde{M}\tilde{a}\right) ,  \label{ZtotPhiRe}
\end{eqnarray}
\begin{eqnarray}
&&z_{_{tot_{2}}}\approx  
-1+\sqrt{\frac{1+z_{0}}{1-z_{0}}}\left( 1 \mp \sqrt{\tilde{M}}
+\right.  \notag \\
&&\left. \frac{3}{2}\tilde{M} \mp \frac{5}{2}\tilde{M}^{3/2} +\frac{27%
}{8}\tilde{M}^{2} + \tilde{M}\tilde{a}\right) ;  \label{ZtotPhiBl}
\end{eqnarray}
if indeed the peculiar redshift is also small, $z_{0}<<1$,\
then these quantities become
\begin{eqnarray}
z_{_{tot_{1}}} &\approx &\pm \tilde{M}^{1/2}+z_{0}+\frac{3}{2}\tilde{M}\pm
\tilde{M}^{1/2}z_{0}  \notag \\
&\pm& \frac{5}{2}\tilde{M}^{3/2}+\frac{3}{2}\tilde{M}z_{0}+\frac{1}{2}%
z_{0}^{2}-\tilde{a}\tilde{M},
\end{eqnarray}%
\begin{eqnarray}
z_{_{tot_{2}}} &\approx &\mp \tilde{M}^{1/2}+z_{0}+\frac{3}{2}\tilde{M}\mp
\tilde{M}^{1/2}z_{0}  \notag \\
&\mp& \frac{5}{2}\tilde{M}^{3/2}+\frac{3}{2}\tilde{M}z_{0}+\frac{1}{2}%
z_{0}^{2}+\tilde{a}\tilde{M},
\end{eqnarray}%
which reduce to those for the Schwarzschild black hole for $\tilde{a}%
=0$ \cite{nhalclcAPJL2021}, as it should be. 

If we consider a real astrophysical system like the set of megamasers circularly 
orbiting supermassive black holes in the center of active galactic nuclei,
from these relations, we
observe that the leading term in this expansion corresponds to the frequency
shift due to the rotational motion of the probe particles orbiting the black
hole, the so-called Doppler or kinematic shifts. This is the item that
corresponds to the purely Newtonian approach that describes their rotational
motion around a black hole. The subleading term corresponds to the peculiar
motion of the black hole from/toward us as a whole entity. The third item
corresponds to the main (non-rotating) contribution of the gravitational
redshift and constitutes a purely general relativistic effect produced by
the black hole mass. The fourth item in this expansion corresponds to a
special relativistic correction that involves the product of the kinematic
frequency shift and the peculiar redshift. Finally, the spin parameter makes
its appearance just in the eighth term of this series expansion, making it
clear that it encodes a very subtle effect.

Interestingly, we can also find explicit relations for the mass and angular
rotation parameter of the Kerr black hole in terms of $z_{_{tot_{1}}}$, $%
z_{_{tot_{2}}}$, $r_{e}$ and $z_{0}$\ by employing (\ref{TotRed})\ and (\ref%
{TotBlue}). Thus, we use the relations (\ref{TotRed})\ and (\ref{TotBlue})
to obtain%
\begin{equation}
\hat{R}\hat{B}=\frac{1+z_{0}}{\left( 1-2\tilde{M}\right) \left(
1-z_{0}\right) },  \label{RhatBhat}
\end{equation}%
and%
\begin{equation}
\frac{\hat{R}}{\hat{B}}=\frac{1+\tilde{M}^{1/2}\left( \tilde{a}-2\tilde{M}%
^{1/2}+\sqrt{1+\tilde{a}^{2}-2\tilde{M}}\right) }{1+\tilde{M}^{1/2}\left(
\tilde{a}-2\tilde{M}^{1/2}-\sqrt{1+\tilde{a}^{2}-2\tilde{M}}\right) },
\label{RhatOverBhat}
\end{equation}%
where $\hat{R}=1+z_{_{tot_{1}}}$ and $\hat{B}=1+z_{_{tot_{2}}}$. Eq. (\ref%
{RhatBhat}) leads to the following relation for the mass%
\begin{equation}
M=\frac{\hat{R}\hat{B}\left( 1-z_{0}\right) -\left( 1+z_{0}\right) }{2\hat{R}%
\hat{B}\left( 1-z_{0}\right) }r_{e}.  \label{massRhat}
\end{equation}

Now, by replacing this quantity into (\ref{RhatOverBhat}), we obtain the
angular rotation parameter as well%
\begin{equation}
a=\frac{\left( \hat{R}-\hat{B}\right) ^{2}\left( 1+z_{0}\right) -\left( \hat{%
R}+\hat{B}\right) G}{\left( 2\hat{R}\hat{B}\right) ^{3/2}\sqrt{\hat{R}\hat{B}%
\left( 1-z_{0}\right) ^{2}-\left( 1-z_{0}^{2}\right) }}r_{e},
\label{AngularRhat}
\end{equation}%
where $G=\sqrt{\left( \hat{R}^{2}+\hat{B}^{2}\right) \left( 1+z_{0}\right)
^{2}-2\hat{R}^{2}\hat{B}^{2}\left( 1-z_{0}^{2}\right) }$. Note that these
relations\ reduce to (\ref{mass})\ and (\ref{AngularAB}) in the limit $%
z_{0}=0$, as we expected.%

Thus, we have obtained closed formulas for determining both the black hole
mass and spin parameters from very few observational data: the redshift and
blueshift of emitted photons, as well as the orbital radius $r_e$ of the
emitter and the peculiar motion of the black hole encoded in $z_{0}$.

These relations enable us to compute the mass and spin parameters of a black
hole hosted at the core of a galaxy moving with respect to us. Here, it is
worth mentioning that these closed formulas relate the black hole parameters
$M$ and $a$ to the frequency shifts $z_{_{tot_{1}}}$ and $z_{_{tot_{2}}}$,
and the orbital radius $r_{e}$, which are directly measured quantities, as
well as to the peculiar redshift, which is not a measurable quantity, but
can be statistically estimated with the help of relations (\ref{TotRed})-(%
\ref{TotBlue}).

\section{Dependency of the redshift on the azimuthal angle}

\label{z-phi}

In this section, we are going to obtain expressions for the redshift and
blueshift of photons coming from a general point of their orbit in the
equatorial plane. To do so, we should obtain the dependency of the redshift
on the azimuthal angle $\varphi $. The equation of motion of photons ($%
k_{\mu }k^{\mu }=0$) in the equatorial plane is%
\begin{equation}
g_{tt}\left( k^{t}\right) ^{2}+g_{rr}\left( k^{r}\right) ^{2}+2g_{t\varphi
}k^{t}k^{\varphi }+g_{\varphi \varphi }\left( k^{\varphi }\right) ^{2}=0,
\label{PhotonEOM}
\end{equation}%
where $k^{t}$\ and $k^{\varphi }$\ can be found through the Killing vector
fields $\partial _{t}$\ and $\partial _{\varphi }$ and are presented in
relations (\ref{tCompOf4momentum})-(\ref{PhiCompOf4momentum}).

By using (\ref{tCompOf4momentum})\ and (\ref{PhiCompOf4momentum}), the
equation of motion (\ref{PhotonEOM})\ takes the following form%
\begin{equation}
g_{rr}\left( k^{r}\right) ^{2}-\frac{g_{tt}L_{\gamma }^{2}+2g_{t\varphi
}L_{\gamma }E_{\gamma }+g_{\varphi \varphi }E_{\gamma }^{2}}{g_{t\varphi
}^{2}-g_{tt}g_{\varphi \varphi }}=0,
\end{equation}%
that gives $k^{r}$ versus constants of motion and metric components\
\begin{equation}
\left( k^{r}\right) ^{2}=\frac{g_{tt}L_{\gamma }^{2}+2g_{t\varphi }L_{\gamma
}E_{\gamma }+g_{\varphi \varphi }E_{\gamma }^{2}}{g_{rr}\left( g_{t\varphi
}^{2}-g_{tt}g_{\varphi \varphi }\right) }.  \label{rCompOf4momentum}
\end{equation}

Now, we geometrically introduce the auxiliary bidimensional vector $K$
defined by the following decomposition 
\begin{equation}
k^{r}=K\cos \varphi ,  \label{kr}
\end{equation}%
\begin{equation}
rk^{\varphi }=K\sin \varphi ,  \label{kphi}
\end{equation}%
where $K^{2}=\left( k^{r}\right) ^{2}+r^{2}\left( k^{\varphi }\right) ^{2}$%
,\ $\ 0\le\varphi\le2\pi $, and therefore, we can use (\ref%
{PhiCompOf4momentum})\ and (\ref{rCompOf4momentum})\ to obtain $K^{2}$
\begin{equation}
K^{2}=\frac{g_{tt}L_{\gamma }^{2}+2g_{t\varphi }L_{\gamma }E_{\gamma
}+g_{\varphi \varphi }E_{\gamma }^{2}}{g_{rr}\left( g_{t\varphi
}^{2}-g_{tt}g_{\varphi \varphi }\right) }+r^{2}\frac{\left( E_{\gamma
}g_{t\varphi }+L_{\gamma }g_{tt}\right) ^{2}}{\left( g_{t\varphi
}^{2}-g_{tt}g_{\varphi \varphi }\right) ^{2}}.
\end{equation}

On the other hand, substituting (\ref{kr})\ in (\ref{rCompOf4momentum})
leads to the following relation for $K^{2}$%
\begin{equation}
K^{2}=\frac{g_{tt}L_{\gamma }^{2}+2g_{t\varphi }L_{\gamma }E_{\gamma
}+g_{\varphi \varphi }E_{\gamma }^{2}}{g_{rr}\left( g_{t\varphi
}^{2}-g_{tt}g_{\varphi \varphi }\right) \cos ^{2}\varphi }.
\end{equation}

Equating previous relations gives an equation for the light bending
parameter $b_{\varphi }=L_{\gamma }/E_{\gamma }$\ as below%
\begin{eqnarray}
&&\left( g_{tt}b_{\varphi }^{2}+2g_{t\varphi }b_{\varphi }+g_{\varphi
\varphi }\right) \left( g_{t\varphi }^{2}-g_{tt}g_{\varphi \varphi }\right)
\sin ^{2}\varphi  \notag \\
&&-r^{2}g_{rr}\left( g_{tt}b_{\varphi }+g_{t\varphi }\right) ^{2}\cos
^{2}\varphi =0,
\end{eqnarray}%
that leads to the solution for $b_{\varphi }$
\begin{equation}
b_{\varphi }=-\frac{g_{t\varphi }}{g_{tt}}-\frac{\left( g_{t\varphi
}^{2}-g_{tt}g_{\varphi \varphi }\right) \sin \varphi }{g_{tt}\sqrt{\left(
g_{t\varphi }^{2}-g_{tt}g_{\varphi \varphi }\right) \sin ^{2}\varphi
-r^{2}g_{tt}g_{rr}\cos ^{2}\varphi }},  \label{bGamma}
\end{equation}%
where we should recall that $0\leq \varphi \leq 2\pi $. Note that this
equation plausibly reduces to relation (\ref{LBParameter})\ for $\varphi
=\pm \pi /2$\ and to (\ref{DragBendLigh})\ for $\varphi =0$, as it should be.

This formula for the light bending parameter is quite remarkable since it
unifies the two expressions we had for this quantity when considering the
motion of a particle on either side of the line of sight. We shall see below
that it unifies the relations (\ref{KerrRedFarDetector})-(\ref%
{KerrBlueFarDetector}) for the Kerr redshift into a single one as well.
Besides, it is a relation encoding the light bending for an arbitrary point 
of the orbit on the equatorial plane.

Since we are investigating photons traveling on the equatorial plane, the
relation (\ref{bGamma}) takes the following explicit form%
\begin{eqnarray}
&&b_{\varphi }=-\frac{2aM}{r-2M}  \notag \\
&&+ \frac{r\Delta ^{3/2}\sin \varphi }{\left( r-2M\right) \sqrt{\Delta
^{2}\sin ^{2}\varphi +\left( r-2M\right) r^{3}\cos ^{2}\varphi }},
\label{bVsPhiMa}
\end{eqnarray}%
which generalizes the known expressions for the light bending parameter (\ref%
{LBParameter})\ and (\ref{DragBendLigh})\ to an arbitrary value of the
azimuthal angle $\varphi $ along the circular orbit of a particle in
geodesic motion around a Kerr black hole.

Here, it is evident that the light bending parameter $b_{\varphi }$ does not
vanish on the line of sight where $\varphi =0$ due to the dragging effect
produced by the spin of the black hole. This fact implies that giving this
parameter an impact factor interpretation necessarily neglects the dragging
effect and therefore is misleading.

\begin{figure*}[tbp]
\centering
\includegraphics[width=0.4\textwidth]{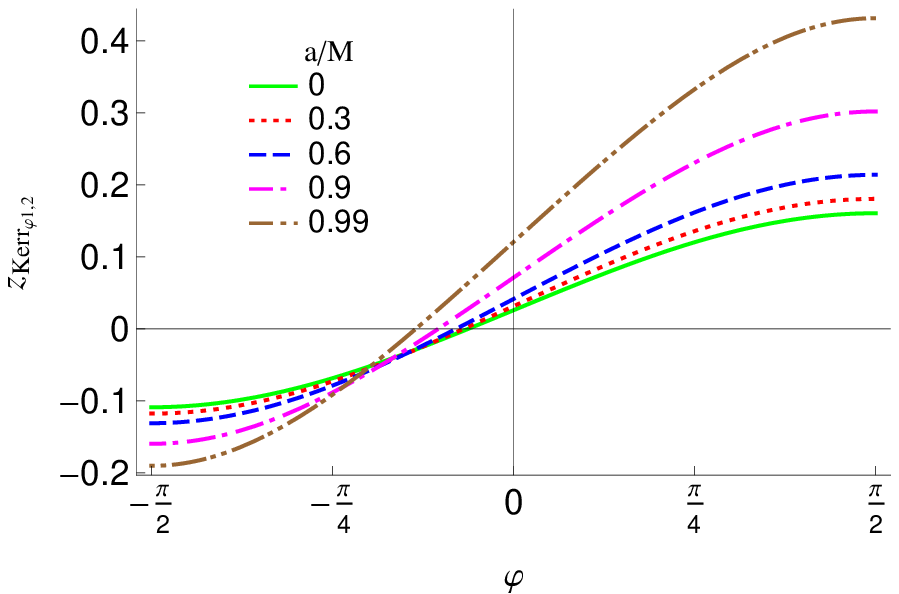}
\includegraphics[width=0.4\textwidth]{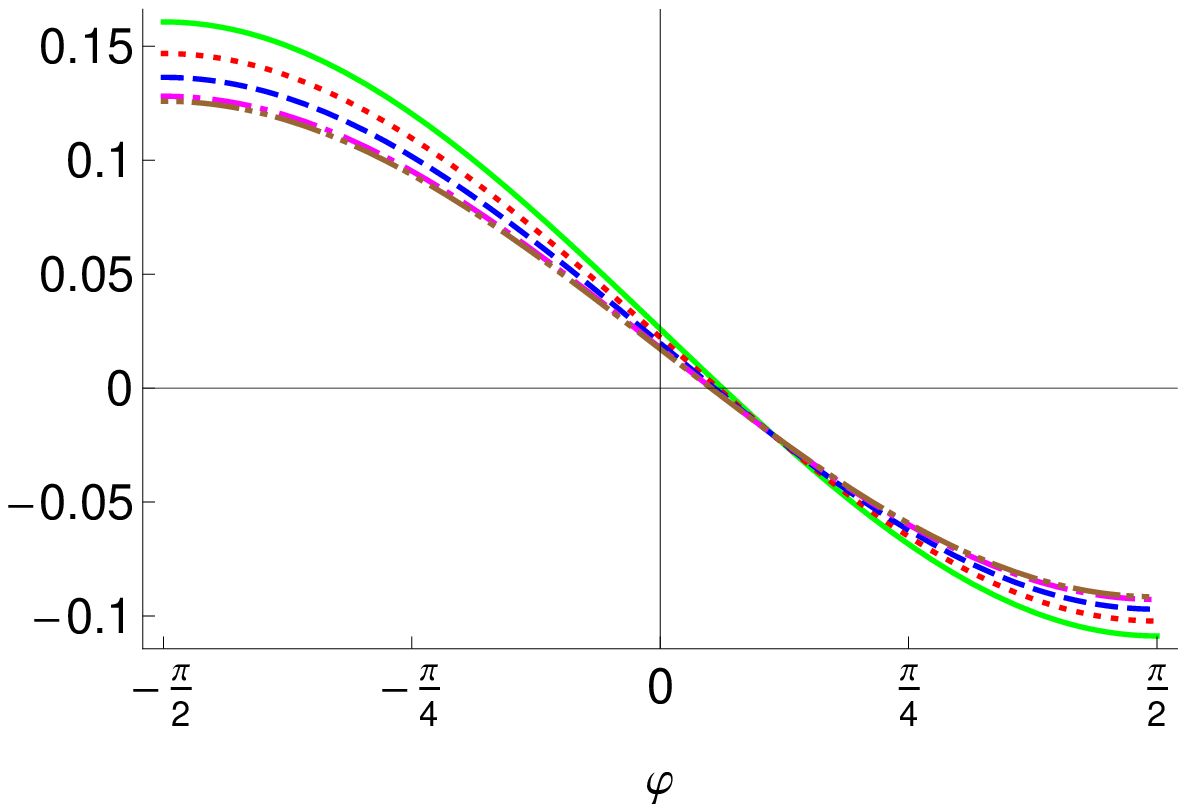}
\caption{The $\protect\varphi $-dependent frequency shifts $z_{_{Kerr_{%
\protect\varphi \,1,2}}}$ (\protect\ref{zphi}) versus the
azimuthal angle for the co-rotating branch (left panel) and
counter-rotating branch (right panel). The continuous green curves
represent the Schwarzschild redshift/blueshift. This figure
is evaluated on $r_{e}=10r_{ms}$.} \label{KerrPhiFig}
\end{figure*}

Having the light bending $b_{\varphi }$ given in (\ref{bVsPhiMa}), we look
for redshift expressions as the next step. For a far away detector with $%
r_{d}>>M\geq a$, the frequency shift (\ref{zcircorbits}) reduces to
\begin{equation}
1+z_{_{Kerr_{\varphi \,1,2}}}=U_{e}^{t}-b_{\varphi }\,U_{e}^{\varphi },
\notag
\end{equation}%
where we used the fact that $U_{d}^{\mu }=\delta _{t}^{\mu }$, a relation obtained
through Eqs. (\ref{tCompVelo}) and (\ref{phiCompVelo}) in the limit $%
r\rightarrow \infty $. We replace $U_{e}^{t}$\ and $U_{e}^{\varphi }$\ from (%
\ref{tCompVelo})\ and (\ref{phiCompVelo})\ to find%
\begin{equation}
1+z_{_{Kerr_{\varphi \,1,2}}}=\frac{\pm M^{1/2}\left( a-b_{\varphi }\right)
+r_{e}^{3/2}}{r_{e}^{3/4}\sqrt{r_{e}^{3/2}-3Mr_{e}^{1/2}\pm 2aM^{1/2}}}.
\end{equation}%

Now, by substituting $b_{\varphi }$\ from (\ref{bVsPhiMa}), we obtain the
following explicit form for the frequency shift%
\begin{eqnarray}
\label{zphi} 
&&1+z_{_{Kerr_{\varphi \,1,2}}}\!\!=\!  
\frac{1}{\left( 1\!-\!2\tilde{M} \right) \sqrt{1\!-\!3\tilde{M}\pm 2\,\tilde{a}\,\tilde{M}^{1/2}}}\times  
\\
&&\left[ 1\!-\!2\tilde{M} \!\pm\!\tilde{M}^{1/2}\!\!\left( \!\tilde{a}+
\frac{\tilde{\Delta}_{e}^{3/2}\sin \varphi }{\sqrt{\tilde{\Delta}
_{e}^{2}\sin ^{2}\!\varphi \!+\!( 1\!-\!2\tilde{M}) \cos ^{2}\!\varphi }}%
\right) \!\right]\!,  \notag
\end{eqnarray}%

for an arbitrary point of the orbit on the equatorial plane that reproduces the expressions for the redshift (\ref%
{KerrRedFarDetector}) and the blueshift (\ref{KerrBlueFarDetector}) when $%
\varphi =\pm \pi /2$, respectively, as well as the formula (\ref%
{CentralRedshift}) for the gravitational redshift when the azimuthal angle
vanishes $\varphi =0$.

When one tries to solve for the mass parameter by making use of the product $%
(1+z_{_{Kerr_{\varphi \,1}}})(1+z_{_{Kerr_{\varphi \,2}}})$ corresponding to
probe particles lying at the same angular distance to the left and to the
right of the line of sight, an eight order algebraic equation for $M$ arises.

Figure \ref{KerrPhiFig} shows the general behavior of $z_{_{Kerr_{\varphi
\,1,2}}}$\ versus the azimuthal angle $\varphi $. For the Schwarzschild
black hole, the continuous green curve confirms that it does not matter
whether the geodesic particle is co-rotating or counter-rotating. Besides, $%
\left\vert z_{_{Kerr_{\varphi \,1,2}}}\right\vert $ for $\left\vert \varphi
\right\vert >\pi /4$\ is an increasing function of $a$ for co-rotating
branch, whereas it is a decreasing function for counter-rotating particles.
This uncovers the importance of particles' angular momentum orientation in
this analysis and shows that the behavior of $z_{_{Kerr_{\,1,2}}}$ for $%
r_{e}=r_{ms}$\ (dashed curves in Fig. \ref{KerrShiftFig}) is valid for $%
r_{e}>r_{ms}$\ as well.

There are also two notable points. First, from the left panel of Fig. \ref%
{KerrPhiFig}, we see non-zero $z_{_{Kerr_{\varphi \,1,2}}}$ at the line of
sight ($\varphi =0$) for the Schwarzschild solutions which increases for the
Kerr solutions as the rotation parameter increases. The non-zero $%
z_{_{Kerr_{\varphi \,1,2}}}$ for the Schwarzschild case is due to the
gravitational redshift (Eq. (\ref{CentralRedshift}) as $\tilde{a}\rightarrow
0$), whereas its upward moving for the Kerr black holes is because of the
dragging effect encoded in Eq. (\ref{CentralRedshift}) due to (\ref%
{DragBendLigh}). Second, we observe vanishing $z_{_{Kerr_{\varphi \,1,2}}}$\
at a critical angle, say $\varphi $$=\bar{\varphi}<0$, which means the
kinematic blueshift cancel the gravitational redshift (we have $z_{_{kin_{%
\mathbf{\bar{\varphi}}-}}}=-z_{g_{\mathbf{\bar{\varphi}}}}$ at $\bar{\varphi}
$, hence $z_{_{Kerr_{\mathbf{\bar{\varphi}}\,1,2}}}=z_{_{kin_{\mathbf{\bar{%
\varphi}}-}}}+z_{g_{\mathbf{\bar{\varphi}}}}=0$). The position of
this root moves to the left by increasing the rotation parameter
due to the dragging effect (Note: the dragging effect increases
the redshift of particles in this case). However, one can see the
opposite behavior for the counter-rotating branch in the right
panel of Fig. \ref{KerrPhiFig}. These two points are among the
crucial findings of the present study.

For astrophysical applications it is important as well to compute the
redshift for bodies lying in the vicinity of the line of sight on their
orbital motion. Thus, for angles close to zero (either side of the line of
sight where $\varphi \approx 0$), we have%
\begin{eqnarray}
&&1+z_{_{Kerr_{\varphi \,1,2}}} \approx  \label{zKerrphismall} \\
&&\frac{\left( 1-2\tilde{M}\right) \pm \tilde{M}^{1/2}\left[ \tilde{a}%
+\left( 1-2\tilde{M}\right) ^{-1/2}\tilde{\Delta} _{e}^{3/2}\varphi \right]
}{\left( 1-2\tilde{M}\right) \sqrt{1-3\tilde{M}\pm 2\tilde{M}^{1/2}\tilde{a}}%
},  \notag
\end{eqnarray}%
where it is worth noticing that the angle $\varphi $ is negative if measured
clockwise with respect to the line of sight. %

This expression for orbiting objects with large radius ($r_{e}>>M,a$) reduces to
\begin{equation}
z_{_{Kerr_{\varphi \,1,2}}}\approx \frac{3}{2}\tilde{M}
\pm \sqrt{\tilde{M}}\varphi +\frac{27}{8}\tilde{M}^{2}\pm \tilde{M}^{3/2}\left(\frac{3}{2}\varphi  -
\tilde{a}\right).  \label{SmallPhiLargeRe}
\end{equation}%

From these relations, we have%
\begin{equation}
\varpi \equiv z_{_{Kerr_{\varphi \,1}}}-\ z_{_{Kerr_{\varphi \,2}}}=\pm  \sqrt{\tilde{M}}\left( 2+3\tilde{M}%
\right) \varphi ,
\end{equation}%
\begin{equation}
\lambda \equiv z_{_{Kerr_{\varphi \,1}}}+\ z_{_{Kerr_{\varphi \,2}}}=\tilde{M}\left( 3 \mp 2%
\tilde{a}\sqrt{\tilde{M}}+\frac{27}{4}\tilde{M}\right)  ,
\end{equation}%
that lead to the following approximate expressions for $M$\ and $a$\ in
terms of the redshift and blueshift
\begin{equation}
M\approx \frac{r_{e}}{9\Gamma }\left( \Gamma -2\right) ^{2},  \label{Mapprox}
\end{equation}%
\begin{equation}
a\approx \frac{9\left[ \left( \Gamma -2\right) ^{2}\left( \Gamma
^{2}+4\right) -12\lambda \Gamma ^{2}\right] r_{e}}{8\sqrt{\Gamma }\left(
\Gamma -2\right) ^{3}},  \label{aapprox}
\end{equation}%
with%
\begin{equation}
\Gamma =\left[ 8+\frac{9\varpi }{2\varphi ^{2}}\left( 9\varpi +\sqrt{%
81\varpi ^{2}+32\varphi ^{2}}\right) \right] ^{1/3}.  \label{Gamma}
\end{equation}

Here, we should note that these expressions indicate approximate closed formulas for the
black hole parameters $M$ and $a$ in terms of observable redshift and
blueshift of photons emitted in the vicinity of the line of sight whereas
the similar exact expressions given in (\ref{mass}) and (\ref{AngularAB}) were
obtained for either side of the black hole, where the position vector of
orbiting object with respect to the black hole location is orthogonal to the
line of sight. Therefore, the relations (\ref{mass}), (\ref{AngularAB}), (%
\ref{Mapprox}), and (\ref{aapprox}) indicate closed formulas for $M$ and $a$
versus $z_{_{Kerr_{\varphi \,1,2}}}$\ that can be considered for a single
orbit but at different emission points, namely, $\varphi =\pm \pi /2$\ and $%
\varphi \approx 0$. The expressions (\ref{Mapprox})-(\ref{aapprox}) can find
astrophysical applications when modeling the frequency shift of photons
emitted by systemic water masers located on accretion disks revolving around
a black hole at the core of active galactic nuclei (see below).

\subsection{The Schwarzschild black hole mass {\it versus} the azimuthal
angle $\protect\varphi$}

Even though Eq. (\ref{zphi}) cannot be algebraically solved for the Kerr
black hole parameters $M$ and $a$, it renders a closed formula for the black
hole mass in the Schwarzschild case when the spin parameter is neglected%
\begin{equation}
M=\frac{H_{+}+\sqrt{H_{-}^{2}+6\mathcal{RB}
\sin ^{2}(2\varphi) }}{12%
\mathcal{RB}\sin ^{2}\varphi }r_{e},
\end{equation}%
with $H_{\pm}=3\left(\mathcal{RB}-\sin ^{2}\!\varphi \right)\pm 2\mathcal{RB}\sin ^{2}\!\varphi $, 
$\,\mathcal{R}%
=1+z_{_{Kerr_{\varphi \,1}}}\!$, and $\mathcal{B}%
=1+z_{_{Kerr_{\varphi \,2}}}$\ which correctly reproduces the expression (%
\ref{mass}) for the mass at the points of maximal emission, i.e., when the
azimuthal angle is equal to $\pm \pi /2$.

It might seem that this formula does not hold for computing the mass
parameter through the redshift measured at the line of sight where $\varphi
=0$. However, in this limiting case one recovers from (\ref{zKerrphismall})
the expression for the central redshift given in (\ref{CentralRedshift}).
For the Schwarzschild case, we find the following formula for the mass
measured at the line of sight
\begin{equation}
M=\frac{\left( 1+z_{g}\right) ^{2}-1}{3\left( 1+z_{g}\right) ^{2}}r_{e},
\end{equation}%
where the gravitational redshift is given by (\ref{CentralRedshift}) when $a$
vanishes.

\section{Boosting the solution with incorporated azimuthal angle}

\label{Boostingz-phi}

At the final stage, it is crucial to obtain the dependency of redshift on
the azimuthal angle $\varphi $ for the black holes experiencing the peculiar
redshift $z_{0}$. This will allow us to include the redshift of emitted
photons from an arbitrary point in a moving galaxy receding from us or
approaching toward us in the calculations. To do so, we compose the $%
\varphi $-dependent redshift $z_{_{Kerr_{\varphi \,1,2}}}$\ given in (\ref%
{zphi}) with the peculiar redshift $z_{0}$, as described in Sec. \ref%
{boosting}.

By considering the relation $z_{_{Kerr_{\varphi \,1,2}}}=z_{_{kin_{\varphi
\pm }}}+z_{g_{\varphi }
}$, the $\varphi $-dependent total redshift $%
z_{_{tot_{\varphi \,1,2}}}$ (\ref{boost})\ takes the form%
\begin{equation}
z_{_{tot_{\varphi \,1,2}}}=\left( 1+z_{_{Kerr_{\varphi \,1,2}}}\right)
\left( 1+z_{boost}\right) -1.
\end{equation}

Now, by employing Eq. (\ref{zphi}), we can find the explicit form of $%
z_{_{tot_{\varphi \,1,2}}}$\ as below%
\begin{eqnarray}
&&z_{_{tot_{\varphi \,1,2}}}=  \notag \\
&&-1+\frac{1}{\left(1-2\tilde{M}\right) \sqrt{1-3\tilde{M}\pm 2\,\tilde{a}\,\tilde{M}^{1/2}}}\times  \notag \\
&&\sqrt{\frac{1+z_{0}}{1-z_{0}}}\left[ 1-2\tilde{M} \pm
\tilde{M}^{1/2}\times \right.  \notag \\
&&\left. \left( \tilde{a}+\frac{\tilde{\Delta}_{e}^{3/2}\sin \varphi }{\sqrt{\tilde{\Delta}
_{e}^{2}\sin ^{2}\varphi +\left( 1-2\tilde{M}\right) \cos ^{2}\varphi }}%
\right) \right] ,  \label{ZtotPhi}
\end{eqnarray}%
which reduces to (\ref{zphi}) for $z_{0}=0$. This expression is the most
general relation for the redshift $z_{_{tot_{\varphi \,1}}}$\ and blueshift $%
z_{_{tot_{\varphi \,2}}}$ of photons emitted by geodesic particles from an
arbitrary point (charachterized by the azimuthal angle $\varphi $) in the
equatorial plane with the emitter radius $r_{e}$ orbiting (either
co-rotating or counter-rotating) the Kerr black hole with the peculiar
redshift $z_{0}$, while the detector is located far away from the source.

If we set $\varphi = \pm \pi/2 $, we get back the boosted redshift expressions corresponding 
to the maximum value of the light bending parameter, given by (\ref{TotRed}) and (\ref{TotBlue}) 
respectively. The expansions in (\ref{ZtotPhiRe}) and (\ref{ZtotPhiBl}) are also valid for sources 
in close proximity to $\varphi \approx \pm \pi/2 $.

The frequency shifts of orbiting particles close to the line of sight, where
$\varphi \approx 0$, simplify to%
\begin{eqnarray}
&&z_{_{tot_{\varphi \,1,2}}} \approx  \notag \\
&&-1+\sqrt{\frac{1+z_{0}}{1-z_{0}}}\times  \notag \\
&&\frac{\left( 1\!-\!2\tilde{M}\right) \pm \tilde{M}^{1/2}\left[ \tilde{a}%
+\left( 1\!-\!2\tilde{M}\right) ^{-1/2}\!\tilde{\Delta} _{e}^{3/2}\varphi \right]
}{\left( 1-2\tilde{M}\right) \sqrt{1-3\tilde{M}\pm 2\tilde{M}^{1/2}\tilde{a}}%
}.	\label{ZtotPhi0}
\end{eqnarray}

Interestingly, it is also possible to obtain closed formulas for the mass
and spin parameters of the Kerr black hole in terms of the total redshift in the
vicinity of the line of sight incorporating the peculiar redshift $z_{0}$. By
expanding the above relation for large emitter radius, we first compute the
generalized form of Eq. (\ref{SmallPhiLargeRe}) as
\begin{eqnarray}
&&z_{_{tot_{\varphi \,1,2}}}\approx  \notag \\
&&-1+\sqrt{\frac{1+z_{0}}{1-z_{0}}}\left[ 1 + \frac{3}{2}\tilde{M}\pm \sqrt{\tilde{M}}\varphi
\right.  \notag \\
&&\left. +\frac{27%
}{8}\tilde{M}^{2} \pm \tilde{M}^{3/2}\left(\frac{3}{2}\varphi - \tilde{a}\right) \right].  \label{ZtotPhiRed}
\end{eqnarray}

Then, we combine these equations to find the following expressions
\begin{equation}
\hat{\varpi}=\pm\sqrt{\frac{1+z_{0}}{1-z_{0}}}\left( 2+3\tilde{M}\right) \sqrt{%
\tilde{M}}\varphi ,
\end{equation}%
\begin{equation}
\hat{\lambda}=-2+\sqrt{\frac{1+z_{0}}{1-z_{0}}}\left[ 2+\tilde{M}\left( 3\mp2%
\tilde{a}\sqrt{\tilde{M}}+\frac{27}{4}\tilde{M}\right) \right] ,
\end{equation}%
where $\hat{\varpi}=z_{_{tot_{\varphi \,1}}}-z_{_{tot_{\varphi \,2}}}$\ and $%
\hat{\lambda}=z_{_{tot_{\varphi \,1}}}+z_{_{tot_{\varphi \,2}}}$. Now, we
take advantage of these expressions to obtain the following approximate
closed formulas for $M$\ and $a$\ in terms of the total redshift and
blueshift
\begin{equation}
M\approx \frac{r_{e}}{9\hat{\Gamma}\left( 1+z_{0}\right) }\left[ \hat{\Gamma}%
-2\left( 1+z_{0}\right) \right] ^{2},  \label{MapproxZ0}
\end{equation}%
\begin{eqnarray}
&&a\approx  \notag \\
&&\frac{9r_{e}}{8\sqrt{\hat{\Gamma}\left( 1+z_{0}\right) }\left[ \hat{\Gamma}%
-2\left( 1+z_{0}\right) \right] ^{3}}\times  \notag \\
&&\left\{ \left[ \hat{\Gamma}-2\left( 1+z_{0}\right) \right] ^{2}\left[ \hat{%
\Gamma}^{2}+4\left( 1+z_{0}\right) ^{2}\right] \right. -  \label{aapproxZ0} \\
&&\left. 12\hat{\Gamma}^{2}\left( 1+z_{0}\right) \sqrt{1-z_{0}^{2}}\left(
\hat{\lambda}+2-2\sqrt{\frac{1+z_{0}}{1-z_{0}}}\right) \right\} ,
\notag
\end{eqnarray}%
\begin{eqnarray*}
&&\hat{\Gamma}= \\
&&\left[ 8\left( 1+z_{0}\right) ^{3}+\frac{9\hat{\varpi}\left(
1+z_{0}\right) ^{2}}{2\varphi ^{2}}\right. \times \\
&&\left. \left( 9\hat{\varpi}\left( 1\!-\!z_{0}\right) +\sqrt{81\hat{\varpi}%
^{2}\left( 1\!-\!z_{0}\right) ^{2}+32\varphi ^{2}\left( 1\!-\!z_{0}^{2}\right) }%
\right) \right] ^{\frac{1}{3}},
\end{eqnarray*}%
which reduce to Eqs. (\ref{Mapprox})-(\ref{Gamma}) in the limit $z_{0}=0$,
respectively. We recall that the closed formulas for the black hole
parameters $M$ and $a$ in terms of observable total redshift and blueshift
of photons presented in (\ref{massRhat}), (\ref{AngularRhat}), (\ref{MapproxZ0}),
and (\ref{aapproxZ0}) can be employed for a single orbit of radius $r_{e}$
at different emission points $\varphi =\pm \pi /2$\ and $\varphi \approx 0$. 

In particular, the above equations are important to obtain the mass and rotation parameters 
of a Kerr black hole using the detected redshift and blueshift of photons emitted from 
sources which lie on both sides of the line of sight at equal but small angular separations,
a setup of photon sources that can be applied to systemic masers in active galactic nuclei
(see below).

Besides, the total frequency shifts (\ref{ZtotPhiRed}) for low peculiar
redshifts ($z_{0}<<1$) and large orbits of the emitter ($M,a<<r_{e}$) read%
\begin{eqnarray}
z_{_{tot_{\varphi \,1,2}}} &\approx & \frac{3}{2%
}\tilde{M} + z_0 \pm \sqrt{\tilde{M}}\varphi +\frac{27}{8}\tilde{M}^{2} +\frac{3}{2%
}\tilde{M}z_0
\notag \\
&&
+ \frac{1}{2}z_{0}^{2}\pm \frac{3}{2}\tilde{M}^{3/2}\varphi  \mp \tilde{M}^{3/2}\tilde{a}.
\end{eqnarray}


\section{Discussion and final remarks}

In this work, we have solved an inverse problem in order to obtain closed
formulas for the Kerr black hole mass and spin parameters in terms of
directly observable quantities. This analysis has been performed for both 
the cases where the Kerr black holes are static and moving (either 
approaching or receding) with respect to us.

These expressions can be applied to real astrophysical systems where 
stars or water maser clouds, for instance, are equatorially revolving in 
circular motion around a central black hole \cite{Moran,MCP1,MCP3,MCP9}. 
The motion of these so-called megamasers has been tracked for several
galaxies using astrometry techniques to determine positions and spectroscopy
to measure frequency shifts.

We would like to highlight that a simplified version of this formalism for static 
(Schwarzschild) black holes has been already applied to the accretion disk 
with water masers that orbit circularly and equatorially around a supermassive 
black hole hosted at the center of the active galactic nucleus of NGC 4258 
\cite{nhalclcAPJL2021}. In the aforementioned study, the mass-to-distance 
ratio was estimated and, moreover, the gravitational redshift of the closest 
maser to the black hole was quantified using this general relativistic method.
Remarkably, the first estimation of the mass-to-distance ratio for the central 
black hole harbored at the core of the so-called gigamaser TXS-2226-184 
has been also performed in \cite{VGHNAA2022}, where a quantitative 
estimate of  the gravitational redshift of the closest maser cloud to the black 
hole was given as well.

The new findings presented here will also be useful for such real astrophysical 
systems. In particular, the results obtained in Sec. \ref{boosting} are relevant 
for applications to certain astrophysical systems that present the geometrical 
properties as well as the receding motion required for our modeling.
As an example we can mention some supermassive black holes hosted at the
center of active galactic nuclei like NGC 4258 and NGC 2273. Both of these
galaxies possess an accretion disk with water masers that orbit circularly
and equatorially around a black hole center; moreover, they are moving away
from us and their motion is parameterized by the so-called peculiar redshift
$z_0$.

On the other hand, the results presented in Secs. \ref{z-phi} and \ref%
{Boostingz-phi} can be applied to some astrophysical systems that present
circularly orbiting bodies that lie close to the line of sight as in almost
all the water megamaser clouds of accretion disks revolving around a black
hole in the center of active galactic nuclei (see, for instance, \cite{Moran,MCP3}). 
These central systemic masers lie 
precisely around the line of sight and so far have been omitted when fitting 
observational data within a general relativistic modeling, even for the 
Schwarzschild black hole case \cite{nhalclcAPJL2021}.

Besides, the aforementioned expressions for the frequency shifts allow us to
statistically estimate the Kerr black hole parameters $M$ and $a$, as well as the
peculiar redshift $z_{0}$, by making use of a Bayesian fit. Here it is worth
mentioning that in order to attain a physically viable estimation for the
spin parameter, it is necessary to measure all the observable quantities,
the total frequency shifts as well as the emitter positions in the sky, with
enough precision since this parameter introduces a very subtle effect on the
gravitational field generated by the black hole.

Finally, we would like to mention that the aforementioned predicted values of the total redshift and blueshift are bounded for the Kerr black hole metric, placing a simple test on general relativity in its strong field regime. Thus, if the observed values of redshifts and blueshifts of photons emitted by astrophysical sources do not lie within the physically allowed shaded area of Fig. \ref{KerrShiftFig}, as predicted by the Kerr spacetime, then this would imply that the photon source is not orbiting a Kerr black hole, but a different metric.


\section*{Acknowledgements}

All authors are grateful to FORDECYT-PRONACES-CONACYT for support under
grant No. CF-MG-2558591; U.N. also acknowledges support under grant CF-140630. 
M.M. is grateful to CONACYT for providing financial assistance through the postdoctoral grant No. 31155.
A.H.-A. and U.N. thank SNI and PROMEP-SEP and were supported by grants VIEP-BUAP No. 122 and CIC-UMSNH, respectively.


\end{document}